\documentclass[aps,12pt]{revtex4}
\usepackage{graphicx,color,array}

\begin{document}
%%%%%%%%%%%%%%%%%%%%%%%%%%%%%%%%%%%%%%%%%%%%%%%%
\title{Quantum Effects in the Conductivity of a Quasi 2D Electron
Gas}
\author{M Levanda$^*$ and V Fleurov$^{1,\dagger}$}
\address{$^1$Raymond and Beverly Sackler Faculty of Exact Sciences,
School of Physics and Astronomy, \\ Tel-Aviv University, Tel-Aviv
69978 Israel. }
%%%%%%%%%%%%%%%%%%%%%%%%%%%%%%%%%%%%%%%%%%%%%%%%

%%%%%%%%%%%%%%%%%%%%%%%%%%%%%%%%%%%%%%%%%%%%%%%%
\date{\today}

%%%%%%%%%%%%%%%%%%%%%%%%%%%%%%%%%%%%%%%%%%%%%%%%
\begin{abstract}
We consider the role of the third dimension in the conductivity of
a quasi 2D electron gas. If the transverse correlation radius of
the scattering potential is smaller than the width of the channel,
i.e. the width of the transverse electron density distribution,
then scattering to higher levels of the confinement potential
becomes important, which causes a broadening of the current flow
profile. The resulting conductivity is larger than that obtained
from a 2D Boltzmann equation. A magnetic field, parallel to the
driving electric field, effectively competes with the confining
potential and, in the limit of a strong magnetic field, it is the
field, which largely shapes the electron and current profile,
rather than the potential. As a result the current flow profile
increases and a negative longitudinal magnetoresistivity of the
quasi 2D electron gas may be observed.
\end{abstract}

\maketitle

%%%%%%%%%%%%%%%%%%%%%%%%%%%%%%%%%%%%%%%%%%%%%%%%
%\pacs{Valid PACES appear here.
%{\tt$\backslash$\string pacs\{\}} should always be input,
%even if empty.}
%%%%%%%%%%%%%%%%%%%%%%%%%%%%%%%%%%%%%%%%%%%%%%%%
%\narrowtext
%%%%%%%%%%%%%%%%%%%%%%%%%%%%%%%%%%%%%%%%%%%%%%%%

\section{Introduction}

Impurity-limited conductivity of a quasi-two-dimensional electron
gas (quasi-2DEG) is usually calculated by means of a
quasi-classical 2D Boltzmann equation (see, e.g.
\cite{afs82,sk70,cb85,tb88,sh67,s76}) and references therein).
Quantum corrections to the 2D conductivity are assumed to be only
due to the weak localization or interaction mechanisms. The
starting point of the quasi-classical approach is the Hamiltonian
\begin{eqnarray}
\hat H &=& \sum_{\alpha,{\bf k}_2}\left(E_\alpha +{{\bf k}_2^2
\over 2 m}\right) a^\dagger_{\alpha,{\bf k}_2} a_{\alpha,{\bf
k}_2}\nonumber \\ & + & \sum_{\alpha,\beta,{\bf k}_2,{\bf q}_2}
M_{\alpha,\beta}({\bf q}_2) e^{i{\bf q}_2 {\bf r}_2}
a^\dagger_{\alpha,{\bf k}_2 + {\bf q}_2/2} a_{\beta,{\bf k}_2 -
{\bf q}_2/2},\label{1}
\end{eqnarray}
where
\begin{equation}\label{matr1}
M_{\alpha,\beta}({\bf q}_2) = \int d z \Psi^*_\alpha (z) U({\bf
q}_2,z)\Psi_\beta(z).
\end{equation}
Here $U({\bf q}_2, z)= \displaystyle \int d^2r_2 e^{-i{\bf
q}_2{\bf r}_2} U({\bf r}_2,z)$ where $U({\bf r}_2,z)$ is a random
scattering potential; the subscript 2 denotes here and below the
2D in-plane vectors. $E_\alpha$ is the $\alpha$-th eigenenergy of
the quantum well, $V(z)$, which confines the motion of the
electrons in the z direction. $\Psi_\alpha(z)$ is the
corresponding eigenfunction.

In order to derive a quasi-classical Boltzmann equation one
assumes that only the highest occupied level, $\alpha_0$, of the
potential $V(z)$ is of importance and all transitions, both real
and virtual, to lower or higher levels can be discarded. If
$\lambda_z$ characterizes the width of the wave function
$\Psi_{\alpha_0}(z)$, then the characteristic energy interval
between the state $E_{\alpha_0}$ and the neighboring states can be
estimated as $\displaystyle \Delta E_{\alpha_0} \approx
\frac{\hbar^2}{2m\lambda_z^2}$. Then the above assumption for the
real transitions is justified if the channel is narrow enough and
the temperature is low enough, $\Delta E_{\alpha_0} \gg k_BT$.

Virtual transitions to other states may be caused by the random
scattering potential $U({\bf q}_2,z)$. Its off-diagonal matrix
elements (\ref{matr1}) are negligible if the scattering potential
is smooth, i.e. it is characterized by a large correlation radius,
$r_c \gg \lambda_z$. Then the scattering is responsible only for
an in-plane relaxation of the electron momentum of the
non-equilibrium 2DEG. One may then discard any possible
renormalization of the current due to an admixture of states with
$\alpha \neq \alpha_0$ and retain only the diagonal matrix
elements $M_{\alpha_0,\alpha_0}({\bf q})$ in Eq. (\ref{1}). Then a
quasi-classical 2D Boltzmann equation follows straightforwardly.
If one is interested in the profile of the current flow density
along the $z$ coordinate, it coincides in this case with the
electron density profile, determined by the wave function of the
$\alpha_0$ level, i.e., $j(z)\sim \rho_{\alpha 0}(z) =
|\psi_{\alpha_0}(z)|^2$. The width of this profile may be called
width of the channel.

A completely different situation takes place in the opposite
limit, $r_c \ll \lambda_z$, to be called below the quantum limit.
The scattering potential $U({\bf q}_2,z)$ induces strong
transitions to other levels $\alpha \neq \alpha_0$ of the quantum
well. The matrix elements $M_{\alpha,\beta} ({\bf q}_2)$ with
$\alpha,\beta \neq \alpha_0$ cannot be discarded, and the
conventional approach based on a quasi-classical 2D Boltzmann
equation is not applicable. An admixture of other states with the
wave functions, localized in much wider regions than that of
$\psi_{\alpha_0}(z)$, may lead to a much broader $z$ profile of
the current density. Its decay along the $z$ axis is characterized
by a length $b$, which we call effective width of the channel. It
may essentially exceed the width $\lambda_z$.

The conductivity of a quasi-2DEG may be now sensitive to an
external in-plane magnetic field, which may influence the
$\lambda_z$ value and lead to a corresponding increase of the
effective width of the channel, which carries the current flow. As
a result, a negative magnetoresistivity in an in-plane magnetic
field is expected. It is worthwhile to distinguish this mechanism
of the longitudinal magnetoresistivity from the other mechanism
recently proposed in Ref. \cite{dg00}, which considers a 2D system
without any account of the third dimension. The magnetic field
polarizes electron spins and causes a change of the Fermi energy
and, hence, of the scattering time. It is emphasized that the
mechanism, we propose here, does not consider electron spins or
their polarization at all. It is of a crucial importance below
that the system is {\em quasi two-dimensional} rather than really
{\em two-dimensional}.

\section{2D - conductivity}

A calculation of the conductivity of a quasi-2DEG in the quantum
limit $\lambda_z \gg r_c$ cannot be carried out within the
framework of a conventional quasiclassical 2D Boltzmann equation.
A quantum approach is necessary, which takes into account the
renormalization of all relevant quantities due to an admixture of
various levels, $\alpha \neq \alpha_0$, of the quantum well. We
shall see below that it is equivalent to explicitly accounting for
the $z$ dependence of the current density flow. This analysis can
be best carried out using a quantum kinetic equation for the
Wigner function. A detailed discussion of a gauge invariant
derivation and analysis of such an equation is presented in our
papers \cite{fk78,lf94,lf01}. A discussion of quantum kinetic
equations can be found in books \cite{m90,jh97}, which present
also introductions to the diagrammatic technique, proposed
originally by Keldysh \cite{k65}.

We consider here a model, in which calculations can be carried out
analytically. The confinement potential is harmonic, $V(z) =
\displaystyle \frac{m\omega^2 z^2}{2}$ with the frequency
$\omega$; $m$ is the electron mass. We simplify the problem by
assuming that only the ground state of the harmonic oscillator is
occupied, $\alpha_0 = 0$, then $\lambda_z = u \equiv \displaystyle
\sqrt{\frac{\hbar}{m\omega}}$; $u$ is the amplitude of the
zero-point oscillations. We assume also that electrons can move
freely in the xy-plane and are scattered by a potential, which is
on the average homogeneous and isotropic in the whole three
dimensional space. Its fluctuations are characterized by a certain
correlation radius $r_c$. This radius can be neglected as compared
to any scales important for the motion in the xy plane. However,
following the above discussion we shall keep a finite value of
$r_c$ when considering the $z$ motion of the electrons. Then
\begin{equation}\label{2}
\langle U({\bf r}_2, z)U({\bf r'}_2, z')\rangle = \overline{
U^2}\delta({\bf r}_2 - {\bf r'}_2) l(z-z').
\end{equation}
Here $\overline{U^2}$ is the mean square fluctuations of the
scattering potential. If the scattering potential is created by
short-range defects with a concentration $c$, then $\overline{
U^2} = c U_d^2$ where $U_d$ is the scattering potential of a
defect. The correlation function for the fluctuations in the $z$
direction is chosen in the Gaussian form
\begin{equation}\label{corr1}
l(z-z') = \frac{1}{r_c\sqrt{2\pi}} \exp[-(z-z')^2/2r_c^2].
\end{equation}

Calculating various space conditional moments of the quantum
kinetic equation one obtains an infinite set of equations,
providing the so called hydrodynamic formulation of the problem
(for details see \cite{lf01}). We restrict ourselves by the $s$-
scattering in the xy-plane, which implies that the equations for
the kinetic moment are decoupled from the equations for higher
powers of the electron momentum. In order to receive these three
equations we first take Eq. (31) of our paper \cite{lf01},
multiply it by the 3D electron momentum {\bf p}, and integrate its
left and right hand parts over the variables $\varepsilon$ and
{\bf p}. The resulting equation for the $z$ component of the
nonequilibrium part of the conditional moment is trivial with
zeros in both left and right hand sides. It corresponds to the
absence of a current in the $z$ direction. The equations for the
$x$ and $y$ components read
\begin{equation}\label{kineq1}
e{\bf E}_2n_2\rho(Z) = 2\int \frac{d\varepsilon}{2\pi\hbar}\int
\frac{d^2p_2}{(2\pi\hbar)^2} {\bf p_2} \exp\left\{ie \hbar {\bf
E}_2\left[ \overleftarrow{\partial^\varepsilon} \overrightarrow
{\partial^{\bf p}_2} - \overleftarrow{\partial^{\bf p}_2}
\overrightarrow{\partial^\varepsilon} \right]\right\} I_z
\end{equation}
where
\begin{equation}\label{density1}
n_2\rho(Z) = -2i\int \frac{d^4p}{(2\pi\hbar)^4} G^<(p, Z)
\end{equation}
is the electron density, whose $Z$ distribution is determined by a
function $\rho(Z)$ normalized to unity, meaning that $n_2$ is the
in-plane electron density; $Z$ and ${\bf p}$ are Wigner variables
for the gauge invariant Green functions \cite{lf94,lf01};
$$I_z = \int \frac{dp_z}{2\pi\hbar} \exp \left \{\frac{i}{2} \hbar
\left( \overleftarrow{\partial^Z} \overrightarrow{\partial ^{p_z}}
- \overleftarrow{\partial ^{p_z}} \overrightarrow{\partial^Z}
\right) \right\} \times
$$
\begin{equation}\label{coll1}
\mathrm{Tr}\left( \sigma_Z{\mathbf\Sigma }(Z,p_z) {\mathbf
G}(Z,p_z) - {\mathbf G}(Z,p_z) {\mathbf\Sigma} (Z,p_z)
\sigma_Z\right).
\end{equation}
Here the Groenewold \cite{g46} notations are used, according which
left and right arrows over the differential operators
$\partial^\xi \equiv \displaystyle \frac{\partial}{\partial\xi}$
denote the operators, which act either on the left or the right
functions in the product, respectively. ${\mathbf G}(Z,p_{z})$ and
${\mathbf\Sigma }(Z,p_z)$ are matrices of the Keldysh Green
functions and mass operators. To shorten the notations their
dependencies on the variables ${\bf p}_2$ and $\varepsilon$ were
suppressed.

The mass operator accounting for the $s$ (in the xy plane)
scattering by the potential (\ref{2}) reads
\begin{equation}
\Sigma^{<>}(\varepsilon; Z, p_z) =
\frac{\overline{U^2}}{(2\pi\hbar)^3} \int d^2p_2dp'_z G^{<>}( {\bf
p}_2, \varepsilon; Z, p'_z) \tilde l(p_z - p'_z). \label{inter}
\end{equation}
where $\tilde l(p_z)$ is the Fourier transform of the correlation
function (\ref{corr1}). Eq.(\ref{kineq1}) is solved in the linear,
with respect to the driving electric field ${\bf E}_2$,
approximation. Then the $s$ scattering means that the mass
operators (\ref{inter}) are at equilibrium.

It is important now to distinguish the $Z$ and $p_z$ dependencies
of the quantities in Eq.(\ref{kineq1}). We consider the case when
the Fermi level lies in the lowest subband of the confinement
potential $V(Z)$, then at equilibrium
$$ \rho(Z) = \overline{\rho}_0(Z) \equiv \frac{1}{u\sqrt{\pi }}
\exp \left( -\frac{Z^{2}}{u^2}\right).$$
\begin{equation}\label{equil1}
G^{(0)<>}({\bf p}_2,\varepsilon;Z,p_{z}) = G_2^{(0)<>}({\bf
p}_2,\varepsilon) \rho_0(Z,p_z|u),
\end{equation}
where $G_2^{(0)<>}({\bf p}_2,\varepsilon)$ is the equilibrium two
dimensional Green functions, and
\begin{equation}\label{density2}
\rho_0(Z,p_z|u) = 4\pi\hbar u\sqrt{\pi} \exp
\left\{-\frac{u^2p_z^2}{\hbar^2} \right\} \overline{\rho}_0(Z)
\end{equation}
(see discussion of the shape of the Wigner quasi-distribution
function of a harmonic oscillator in \cite{lf01}).

The time $\tau_2$ of the electron scattering in the xy plane at
equilibrium does not depend on the variables $Z$ or ${\bf p}_2$
and is defined as
\begin{equation}\label{scatt}
\frac{\hbar l(0)}{\tau_2} \int \frac{dp_z'}{2\pi\hbar}
\rho_0(Z,p_z') \tilde l(p_z - p_z') = i(\Sigma^{(0)<}(Z, p_z,
\varepsilon) - \Sigma^{(0)>}(Z, p_z, \varepsilon)).
\end{equation}
with $l(0) = 1/(r_c\sqrt{2\pi})$.

From the principle of the detailed balance for the quantum
collision integral \cite{fk78}, we conclude that the
nonequilibrium parts of the two above Green functions coincide,
i.e. $\delta G^>(Z,p_z;{\bf p}_2,\varepsilon) = \delta G^<(Z,p_z;
{\bf p}_2,\varepsilon) \equiv  \delta G(Z,p_z; {\bf
p}_2,\varepsilon)$. However, their $Z$ and $p_z$ dependencies do
not necessarily coincide with those of the equilibrium Green
functions (\ref{equil1}). These dependencies can be found from the
integral equation, which follows from Eq. (\ref{kineq1}). A
Gaussian dependence of the nonequlibrium part of the Green
function solves this equation. We look for a solution in the form
\begin{equation}\label{nonequil1}
\delta G({\bf p}_2,\varepsilon;Z,p_z) = \delta G_2({\bf
p}_2,\varepsilon) \rho_0(Z,p_z|b)
\end{equation}
where $\rho_0(Z,p_z|b)$ is described by Eq. (\ref{density2}), in
which the length $u$ is substituted for an unknown length $b$.
Here $\delta G_2({\bf p}_2,\varepsilon)$ is the (yet unknown)
nonequilibrium correction to the 2D Green functions, Eq.
(\ref{equil1}).

We now introduce all the above assumptions into the collision
integral (\ref{coll1}), and after some tedious but straightforward
calculations arrive at the expression
\begin{equation}\label{coll6}
I_z(Z)  = -i \frac{\tilde r_c}{ub\tau_2} \delta G_2({\bf
p_2,\varepsilon}) \exp\left\{-Z^2\left(\frac{1}{\tilde u^2} -
\frac{\tilde r_c^2}{2 \tilde u^4} \right)\right\}
\end{equation}
where
$$\displaystyle \frac{1}{\tilde u^2} = \frac{1}{u^2} +
\frac{1}{b^2}$$
and
\begin{equation}
\displaystyle \frac{1}{\tilde r_c^2} = \frac{1}{r_c^2} +
\frac{1}{u^2} + \frac{1}{b^2}. \label{rc}
\end{equation}
As a result the kinetic equation (\ref{kineq1}) acquires the form
\begin{equation}\label{kineq2}
e{\bf E}_2n_2\overline{\rho}_0(Z) = 2\int
\frac{d\varepsilon}{2\pi\hbar} \int
\frac{d^2p_2}{(2\pi\hbar)^2}{\bf p}_2I_z(Z).
\end{equation}

Now we compare the $Z$ dependence of the right hand side of the
kinetic equation (\ref{kineq2}), determined by Eq. (\ref{coll6}),
and that of the left hand side, determined by the distribution
$\overline{\rho}_0(Z)$. The requirement that these dependencies
coincide, leads to the value
\begin{equation}\label{corr2}
\frac{1}{b^2} = - \frac{1}{r_c^2} + \sqrt{\frac{1}{r_c^4} +
\frac{1}{u^4}},
\end{equation}
which determines the scale of the $Z$ dependence of the
nonequilibrium Green function (\ref{nonequil1}). $b$ is now the
effective width of the channel which does not necessarily coincide
with the width $u$ of the electron density profile. One can
readily see that $b\approx u$ only in the limit of the large
correlation radius $r_c \gg u$. However, it may become very large,
$b\approx \displaystyle \frac{u^2}{r_c} \gg u$, in the quantum
limit, $r_c \ll u$.

Eq. (\ref{kineq2}) allows one to find directly the first
conditional moment of the nonequilibrium correction to the
2D-Green function
\begin{equation}\label{green2d}
\overline{{\bf\delta G}_2} \equiv \int
\frac{d\varepsilon}{2\pi\hbar} \int
\frac{d^2p_2}{(2\pi\hbar)^2}{\bf p}_2\delta G_2({\bf
p_2,\varepsilon}) = i\frac{en_2\tau_2}{2}\frac{b}{\tilde r_c}{\bf
E}_2
\end{equation}
and, hence, the density of the electric current in the quasi-2DEG
becomes (see Eq. (45) in \cite{lf01})
\begin{equation}\label{current1}
{\bf J}(Z) = -2ie \int\frac{d\varepsilon}{2\pi\hbar}\int
\frac{d^2p_2}{(2\pi\hbar)^2} \frac{{\bf p}_2}{m}  \delta G_2({\bf
p}_2, \varepsilon) \overline{\rho}_0(Z|b) =  \sigma_2
\frac{1}{\sqrt{2\pi}\tilde r_c}
\exp{\left(-\frac{Z^2}{b^2}\right)} {\bf E}_2.
\end{equation}
where $\sigma_2 = \displaystyle\frac{e^2n_2\tau_2}{m}$ is the
2D-Drude conductivity.

Integrating Eq.(\ref{current1}) over $Z$ one gets the conductivity
of the quasi-2DEG, $\sigma = \sigma_2 \displaystyle
\frac{b}{\sqrt{2}\tilde r_c}$, which strongly depends on the ratio
of the width of the channel $u$ and the correlation length $r_c$
of the scattering potential. The Drude formula is applicable, if
only long range fluctuations are characteristic of the scattering
potential, i.e., $r_c \gg u$, then $\displaystyle
\frac{b}{\sqrt{2} \tilde r_c} \rightarrow 1$, and $\sigma =
\sigma_2$. However, in the quantum limit, $r_c \ll u$, when the
scattering potential fluctuates in the range smaller than the
width of the channel, a strong deviation from the Drude formula is
expected. The effective width of the channel, where the current
actually flows, $b\approx \displaystyle \frac{u^2}{r_c^2} \gg u$,
may become much larger than the width $u$ of the electron density
profile. As a result, the conductivity may also become much larger
$\sigma = \sigma_2 \displaystyle \frac{u^2}{r_c^2} \gg \sigma_2$.

\section{Applicability of the linear response approximation}

The above results are obtained within the linear response
approximation. In order to probe the applicability limits of this
approximation, one considers the $Z$ profile of the drift
velocity. It can be found using the conditional moment
(\ref{green2d}) (see Eq. (43) in \cite{lf01})
\begin{equation}\label{drift}
{\bf v}_{drift} =
\frac{1}{\overline{\rho}_0(Z)}\int\frac{d\varepsilon}{2\pi\hbar}\int
\frac{d^2p_2}{(2\pi\hbar)^2} \frac{{\bf p}_2}{m}  \delta G_2({\bf
p}_2, \varepsilon).
\end{equation}
This definition corresponds also to the conventional equation
${\bf J}(Z) = e n_0(Z) {\bf v}_{drift}(Z)$, in which $n_0(Z) =
n_2\overline{\rho}_0(Z)$ is the electron density at equilibrium.

One can readily see that differing $Z$ dependencies of the current
and the electron densities lead to a growth of the drift velocity
with increasing $Z$. This fact indicates that at however small
electric field $E_2$, there are large enough distances $Z$, at
which conditions can be achieved, when the linear approximation in
$E_2$ is violated. The detailed analysis requires extremely
cumbersome and lengthy calculations, which cannot be presented
here. However, the most important conclusions can be made in a
simpler way. We obtain a rough but sufficient (upper limit)
criterion of the applicability of the linear approximation by
requiring that the additional energy acquired by an electron due
to the current flow does not exceed the temperature, i.e.
$v_{drift}p_F \ll k_BT$ where $p_F$ is the Fermi momentum. Hence,
the linear approximation is violated if $Z > Z^*$ where
$${Z^*}^2 = \displaystyle\frac{u^2b^2}{b^2 - u^2} \ln
\left|\frac{e\tau_2p_F}{m}\frac{u}{\tilde r_c}\frac{E_2}{k_B
T}\right|.$$

It is important to emphasize that although this condition is
obtained for the harmonic potential well, considered here, its
meaning is more general. For any potential well exists a value
$Z^*$ so that at larger distances the linear approximation does
not hold. At $Z > Z^*$ one has to consider a nonlinear problem
involving contributions of various inelastic scattering mechanism
(say, electron-phonon interaction). However, without making
detailed calculations one can understand that in this region the
scattering becomes much stronger and the current flow profile is
actually much lower, than that obtained in the linear
approximation, hence, its contribution to the total current can be
neglected. When integrating over $Z$ one may just cut the
integration at $|Z| = Z^*$ which provides reasonable estimates for
the nonlinear corrections to the 2D conductivity,
$$\frac{\delta\sigma}{\sigma} = -
2 \int\limits_{Z^*/b}^\infty e^{- x^2} dx.$$
$Z^*$ increases with the decreasing electric field, and at $Z^*\gg
b$ these corrections are hardly observable.

\section{Influence of an in-plane magnetic field}

It is worthwhile to discuss here the influence of an in-plane
magnetic field ${\bf B}$ on the conductivity. According to the
intuition based on the classical ideas about the electron motion,
we do not expect any influence of an in-plane magnetic field,
especially if it is directed {\em parallel} to the electric field.
Nevertheless, as we demonstrate below, such an influence exists
and a positive longitudinal magnetoresistivity is expected. We
have to start again from Eq. (31) of our paper \cite{lf01} and
carry out the same procedure as described above. First, one can
readily check that Eq. (\ref{kineq1}) retains its form in the case
of ${\bf B} \| {\bf E_2}$ and the magnetic field ${\bf B}$ does
not appear explicitly. However, it influences the shape of the
equilibrium Wigner function \cite{lf01}, which becomes now
\begin{equation}\label{density3}
\rho_0(Z,p_z|u,\nu) = 4\pi \hbar \nu^{-2} \exp \left\{-\frac{
u^2p_z^2}{\hbar^2 \nu} \right\} \exp\left\{- \frac{1}{u^2 \nu^{3}}
\left[z + \frac{p_x}{\hbar}\frac{u^2}{l_B^2}\right]^2\right\}
\end{equation}
where
$$\nu = \left(1+ \frac{u^4}{l_B^4}\right)^{1/2}$$
and $l^2_B = \displaystyle \frac{\hbar c}{eB}$ is the magnetic
length. We use this function in order to repeat the calculations
carried above for the case of $B = 0$. The solution $\delta G({\bf
p}_2,\varepsilon;Z,p_z)$ of the quantum kinetic equation will be
again looked in the form (\ref{nonequil1}) where the function
(\ref{density3}) with $b$ substituted for $u$ is used instead of
$\rho_0(Z,p_z|b)$.

Then we get an equation determining the parameter $b$ as a
function of the magnetic field
$$
\frac{u^2}{b^2(\nu)} = \frac{\nu^3}{2\nu^3-1}\left\{- \frac{\nu^4
- 1}{2\nu^3} - \frac{u^2}{r_c^2} + \sqrt{ \left( \frac{\nu^4 -
1}{2\nu^3} + \frac{u^2}{r_c^2}\right)^2 + \frac{2\nu^3 -1 }{
\nu^6} }\right\}
$$
As a result the density of the magnetic field dependent electric
current (\ref{current1}) in the quasi-2DEG becomes
\begin{equation}\label{current4}
{\bf J}(Z,\nu)  =  \sigma_2 \frac{1}{\sqrt{2\pi }\tilde r_c(\nu)}
\exp \left( -\frac{Z^{2}}{b^2(\nu)}\right){\bf E}_2.
\end{equation}
where
$$
\frac{u^2}{\tilde r_c^2(\nu)} = \frac{u^2}{r_c^2} +
\frac{1}{2}(\nu + \nu ^{-3})\left(1 + \frac{u^2}{b^2}\right).$$
Integrating Eq.(\ref{current4}) over $Z$ the magnetic field
dependent conductivity of the quasi-2DEG is,
$$\sigma(\nu) = \sigma_2 \displaystyle \frac{b(\nu)}{\sqrt{2}\tilde
r_c(\nu)}$$

\begin{figure}[htb]
{\includegraphics[width=13cm]{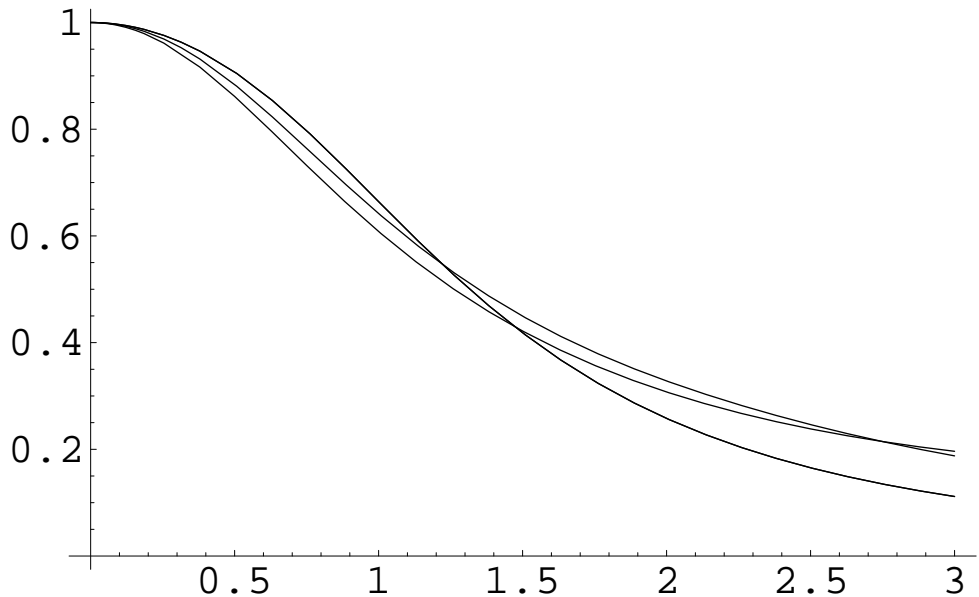} \caption{ Relative 2D
resistivity as a function of the longitudinal magnetic field in
units $u^2/l_B^2$ for different ratios of the parameter $u/r_c$
from 0 to 2. The curves nearly coincide.}}
\end{figure}

Figure 1 shows the relative variation of the resistivity
$\rho(\nu)/\rho(0) = \sigma(0)/\sigma(\nu)$ as a function of the
magnetic field in units $u^2/l^2_B$ for four values of the
parameter $u/r_c$: 0, 0.5, 1, 2. The curves for $u/r_c = 0$ and
0.5 are indistinguishable in this plot. Two other curves for the
higher values of the parameter $u/r_c$ also nearly collapse. In
the high magnetic field limit the resistivity rapidly decreases.
It happens when the magnetic length $l_B$ becomes much smaller
than the width of the channel. Then the dynamics of electrons is
controlled by the magnetic field and the confining potential well
becomes actually irrelevant.

Experimentally this limit can hardly be achieved. If we take a
typical width of the channel to be 100\AA\ then the high magnetic
field limit is reached only in the field measured in hundreds or
even thousands of Tesla. It is much more realistic to expect some
measurements carried out in the low field limit ($l_B > u$) where
a negative magnetoresistivity can be expected. We can write the
low field expansion of the relative magnetoresisitvity as
$$\frac{\rho(B)}{\rho(0)} = 1 - k \frac{u^2}{l_B^2}$$
where
$$
k = \frac{2 + 7 \upsilon^2 + 5 \upsilon^4 + \sqrt{1 + \upsilon^4}
- \upsilon^4(-4 + 5 \sqrt{1 + \upsilon^2})}{4\sqrt{1 + \upsilon^4}
(1 + \sqrt{1 + \upsilon^4}) (- \upsilon^2 + \sqrt{1 +
\upsilon^4})}
$$
The dependence of this coefficient on the parameter $\upsilon^2 =
u^2/r_c^2$ is nearly linear as shown in Figure 2 and starts from
$k(0)= 3/8$ at $\upsilon = 0$. All the above data indicate that
the resistivity may increase by a factor of 2 if the magnetic
length $l_B$ becomes comparable with the width of the channel $u$.
\begin{figure}[htb]
{\includegraphics[width=13cm]{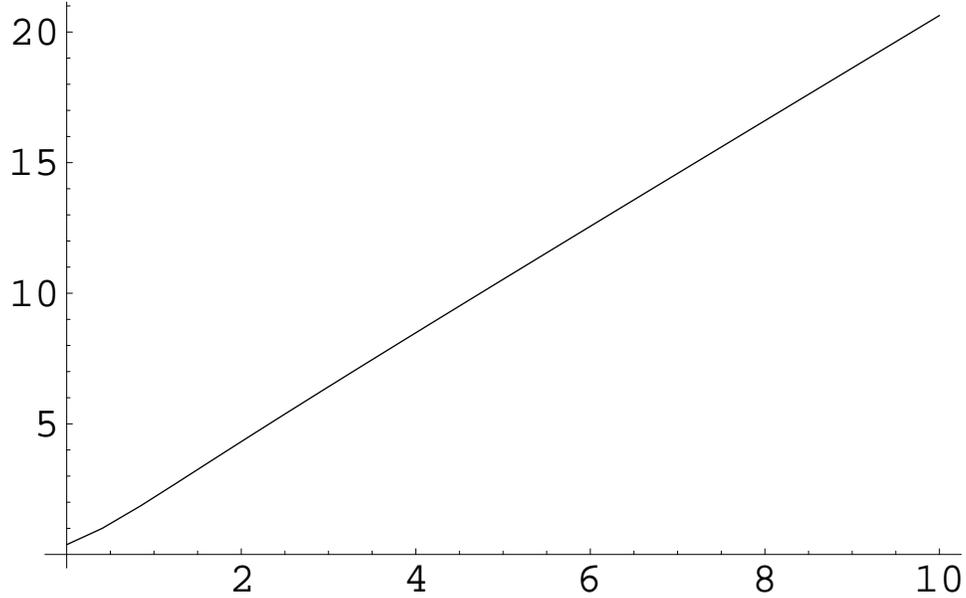} \caption{ The
dependence of the coefficient $k$ in the low field expansion of
the magnetoresistivity on the parameter $\upsilon^2$.}}
\end{figure}

A positive longitudinal magnetoresistivity was observed in SiGe
layers \cite{ohky99,ohkyys99}. The authors attributed the effect
to the electron spin polarization (a theory is presented in
\cite{dg00}). This explanation seems to be quite reasonable since
the electron concentrations in these experiments are so low that a
complete spin polarization is possible. The mechanism proposed in
this paper may become more important at higher electron
concentration. We do not know currently about any relevant
experimental data and the question remains open.

The results of this paper emphasize that the difference between
the 2DEG and quasi-2DEG, however subtle it is, may be of an utter
importance. If the scattering potential comprises fluctuations
with a small enough correlation radius, the profile of the
electron current may become broader than the profile of the
electron density. A longitudinal negative magnetoresistivity is
proposed as an experimentally observable consequence.

\end{document}